\documentclass[12pt]{JHEP3}
\usepackage{amsmath,amssymb,bm}

\author{Shinya Tomizawa$^{\dagger}$, Hideki Ishihara $^{\S}$, Masashi Kimura$^{\ast}$ and Ken Matsuno$^{\ddagger}$\\

Department of Mathematics and Physics,
Graduate School of Science, Osaka City University,
3-3-138 Sugimoto, Sumiyoshi, Osaka 558-8585, Japan\\

$^{\dagger}$\email{tomizawa@sci.osaka-cu.ac.jp},
$^{\S}$\email{ishihara@sci.osaka-cu.ac.jp},\\
$^{\ast}$\email{mkimura@sci.osaka-cu.ac.jp},
$^{\ddagger}$\email{matsuno@sci.osaka-cu.ac.jp}}
\title{Supersymmetric Black Rings on Eguchi-Hanson Space}
\abstract{We construct new supersymmetric black ring solutions on the Eguchi-Hanson base space as solutions of five-dimensional minimal supergravity. The solutions have  the same two angular momentum components and the asymptotic structure on timeslices is asymptotically locally Euclidean. The $\rm S^1$-direction of the black ring is along the equator on a $\rm S^2$-bolt on the Eguchi-Hanson space. We also investigate the limit to a black hole, which describes the BMPV black hole with the topology of the lens space $L(2;1)=\rm S^3/{\mathbb Z}_2$.}

\preprint{OCU-PHYS 266\ AP-GR 42}

\maketitle

\begin{document}


\section{Introduction}

In recent years, the studies on higher dimensional black holes
have attracted much attention in the context of string theory
and the brane world scenario since it is suggested that higher dimensional black holes would be produced in a future linear collider~\cite{BHinCollider}.
Such physical phenomena are expected not only to give us
the proof of the existence of extra dimensions
but also to let us know some information toward quantum gravity. Some of studies on higher-dimensional black holes
show that they have much more complicated and richer structure
than four-dimensional ones.
For an example, in asymptotically flat spacetimes, the topology of the event horizon
in higher dimensions cannot be uniquely determined~\cite{Cai,HelfgottGalloway}
in contrast to four-dimensional ones, which is restricted only to ${\rm S}^2$ under dominant energy condition~\cite{Hawking,hawking_ellis}.
In five dimensions, the possible horizon topology
is either ${\rm S}^3$ or ${\rm S}^1\times {\rm S}^2$~\cite{Cai}. 
In dimensions higher than five, black holes will have more complicated structure~\cite{HelfgottGalloway}.

In fact, asymptotically flat black holes with spherical topology were found by Myers and Perry as a vacuum solution in general dimensional Einstein equations~\cite{Myers}. Emparan and Reall~\cite{Emparan} found the first black ring solution of the five-dimensional vacuum Einstein equation, which describes a stationary black hole rotating along $\rm S^1$-direction with the event horizon homeomorphic to $\rm S^1\times \rm S^2$. In a certain region of parameters, the five-dimensional Myers-Perry black hole with a single angular momentum component and the Emparan-Reall black ring carry the same mass and the angular momentum component, which implies the nonuniqueness theorem in higher dimensional stationary black hole solution. Recently, using one of solitonic solution-generation techniques, i.e.,  the B\"acklund transformation, Mishima and Iguchi found the black ring solution with rotating two sphere, which is regenerated by the another solitonic solution-generation technique~\cite{MI}, i.e., the inverse scattering method by one of authors~\cite{Tomizawa}. Pomeransky et.al.~\cite{Pomeransky}  seems to have found the new black ring solution with two independent angular momentum components by using the inverse scattering method and the result obtained in the references~\cite{Tomizawa2,MI2}. Elvang and Figureas generated a black Saturn solution, which describes a spherical black hole surrounded by a black ring, by the inverse scattering method~\cite{EF}. Furthermore, Iguchi and Mishima also generated the black-di ring solution~\cite{MI3}. 

On the other hand, the discovery of supersymmetric black ring solutions is important in that they preserve supersymmetry and give the BPS (Bogomol'nyi-Prasad-Sommerfield) states. Hence, such supersymmetric black hole/ring solutions are expected to realize the stable state. The supersymmetric solutions of five-dimensional minimal supergravity have been studied by a lot of authors. For example, it was shown that the possible type topology of the cross section of an event horizon is (possibly a quotient of) a homogeneously squashed $S^3$, $S^1\times S^2$ and $T^3$~\cite{Reall}. The only asymptotically flat black hole solution with the topology of $S^3$ is the BMPV(Breckenridge-Myers-Peet-Vafa) solution~\cite{BMPV}, which is specified by two equal angular momenta and the mass~\cite{Reall}. They have been constructed on hyper-K\"ahler base spaces, especially, the Gibbons-Hawking base space. Elvang et.al. found the first supersymmetric black ring solution with asymptotic flatness on the four-dimensional Euclidean base space, which depend on three parameters, mass and two independent angular momentum components~\cite{Elvang}. Gauntlett and Gutowski also constructed a multi-black ring solution on the base space~\cite{Gauntlett,Gauntlett2}. Bena et.al constructed the most three-charge, three-dipole-charge, BPS supersymmetric black ring solution on the Taub-NUT base space~\cite{Bena}. 

A lot of physicists are also specially attracted with black hole solutions with asymptotically Euclidean time slices since they would be a good idealization in the situation such that we can ignore the tension of the brane and the curvature radius of the bulk, or the size of extra dimensions. 
However, from more realistic view point, we need not impose the asymptotic Euclidean condition toward the extra dimensions. 
In fact, higher dimensional black holes admit a variety of asymptotic structures: Kaluza-Klein black hole solutions~\cite{IM,IKMT} have the spatial infinity 
with compact extra dimensions; Black hole solutions on the Eguchi-Hanson space~\cite{IKMT2} have the spatial infinity of topologically various lens spaces $L(2n;1)={\rm S}^3/{\mathbb Z}_{2n}$ ($n$:natural number). Since the latter black hole spacetimes are asymptotically locally Euclidean, we cannot locally distinguish these asymptotic structure.
In spacetimes with such asymptotic structures, furthermore, 
black holes have the structures considerably different from the black hole 
with the asymptotically Euclidean structure. 
For instance, the Kaluza-Klein black holes~\cite{IM,IKMT} 
and the black holes on the Eguchi-Hanson space~\cite{IKMT2} can have the horizon of lens spaces in addition to ${\rm S}^3$. 

As solutions in five-dimensional Einstein-Maxwell theory with a positive cosmological constant, black hole solutions on Taub-NUT space~\cite{IIKMMT} and Eguchi-Hason space~\cite{IMK} were also constructed by the present authors. These multi-black hole solutions describes the non-trivial coalescence of black holes, which is brought about by the non-trivial asymptotic structure. In the reference~\cite{IMK}, we investigated how the coalescence of five-dimensional two black holes depends on the asymptotic structure of spacetime and compared with the five-dimensional Kastor-Traschen solution. Namely, two black holes with the topology of $\rm S^3$ coalesce into a single black hole with the topology of the lens space $L(2;1)=\rm S^3/{\mathbb Z}_2$, while in the Kastor-Traschen solution, two black holes with the topology of $\rm S^3$ coalesce into a single black hole with the topology of $S^3$. The difference helps us know what kind of asymptotic structure we live in the world with. This is why we need to study a black hole solution with non-trivial asymptotic structure.

Our end of this article is to construct new supersymmetric stationary black ring solutions on the Eguchi-Hanson space which has asymptotically locally Euclidean timeslices as solutions of five-dimensional minimal supergravity. The black ring has two equal angular momentum components in two orthogonal planes. Interestingly, the $\rm S^1$-direction of the black ring must be along the equator on a $\rm S^2$-bolt since, otherwise, a Dirac-Misner string would arise near the nuts on the $\rm S^2$-bolt. We also investigate the limit of the black ring to black hole. Finally, we discuss multi-black ring solutions on the Eguchi-Hanson space and comment on the configuration of black rings.

The rest of this article is organized as follows. First, we partially review the results of classification of solutions of five-dimensional minimal supergravity in section \ref{sec:pre}. In section \ref{sec:solutions}, we construct a new supersymmetric black ring solution on the Eguchi-Hanson space. In section \ref{sec:properties}, we study the properties of our solutions, especially, the near-horizon geometry and the black hole limit. We conclude our article with a discussion in section \ref{sec:summary}.


\section{Preliminaries }\label{sec:pre}
The bosonic sector of five-dimensional minimal supergravity is Einstein-Maxwell theory with a Chern-Simon term. Following the reference \cite{Gauntlett3},
all supersymmetric solutions of five-dimensional minimal supergravity have a non-spacelike Killing vector field. In a region where the Killing vector field $\partial/\partial t$ is timelike,
the metric and the gauge potential are given by
\begin{eqnarray}
ds^2=-H^{-2}(dt+\bm\omega)^2+Hds^2_{{\cal B}},\ {\bm A}=\frac{\sqrt{3}}{2}[H^{-1}(dt+\bm\omega)-\bm\beta],
\end{eqnarray}
respectively, where $ds^2_{\cal B}$ is a metric of a hyper-K\"ahler space ${\cal B}$. The scalar function $H$, one-forms $\bm\omega$ and $\bm\beta$ on ${\cal B}$ are given by 
\begin{eqnarray}
\Delta H=\frac{4}{9}(G^+)^2,\quad dG^+=0,\quad d{\bm \beta}=\frac{2}{3}G^+.
\end{eqnarray}
Here, $\triangle$ is the Laplacian on ${\cal B}$ and the two-form $G^+$ is the self-dual part of the one-form $H^{-1}{\bm \omega}$, which is given by 
\begin{eqnarray}
G^+:=\frac{1}{2}H^{-1}(d\omega+*d\omega).
\end{eqnarray}
$(G^+)^2:=\frac{1}{2}G_{mn}G^{mn}$ and $*$ is the Hodge dual operator on ${\cal B}$. 
Since $\partial/\partial t$ is a Killing vector field associated with time translation, all components does not depend on the time coordinate $t$. In this article, as the base space ${\cal B}$, we consider the Gibbons-Hawking metric, which is given by
\begin{eqnarray}
ds^2_{\cal B}=H_k(dr^2+r^2d\Omega_{S^2}^2)+H_k^{-1}(d\xi+\bm\varphi)^2,
\end{eqnarray}
where $H_k$ is a harmonic function on the three-dimensional Euclid space ${\mathbb E}^3$.
Here, $\bm\varphi$ is determined by ${\rm rot}\ \bm\varphi={\rm grad}\ H_k$, and $\partial/\partial \xi$ is a Killig vectors with closed orbits on the base space. If the Killing vector $\partial/\partial \xi$ is also a Killing vector field of the full five-dimensional spacetime, i.e., if $H$ and ${\bm \omega}$ are independent of $\xi$, $H$ and ${\bm \omega}$ can be solved explicitly. The one-forms $\bm\omega$ and $\bm\beta$ can be written in the form
\begin{eqnarray}
\bm\beta&=&\beta_0(d\xi+\bm\varphi)+\tilde{\bm\beta}, \\
 \bm \omega&=&\omega_0(d\xi+\bm\varphi)+\tilde{\bm\omega},
\end{eqnarray}
The functions $H$, $\omega_0$ and $\beta_0$ are written as 
\begin{eqnarray}
       H&=&H_k^{-1}K^2+L,\\
       \beta_0&=&H_k^{-1}K,\\
\omega_0&=&H_k^{-2}K^3+\frac{3}{2}H_k^{-1}KL+M,
\end{eqnarray}
where $K$, $L$ and $M$ are other harmonic functions on ${\mathbb E}^3$. The one-forms $\tilde{\bm\omega}$ and $\tilde{\bm\beta}$ are determined by the equations
\begin{eqnarray}
d{\tilde{\bm \omega}}=* \left[ H_{k}dM-MdH_{k}+\frac{3}{2}(KdL-LdK)\right],
\end{eqnarray}
\begin{eqnarray}
d\tilde{\bm\beta}=-*dK,
\end{eqnarray}
where $*$ denotes the Hodge dual on a three dimensional Euclid space $ds^2=dr^2+r^2(d\theta^2+\sin^2\theta d\phi^2)$. In this article, we restrict the form of $H_k$ to  
\begin{eqnarray}
H_k&=&\epsilon+\sum_i\frac{Q_k^{(i)}}{\Delta_{a_i}},
\end{eqnarray}
where $Q_k^{(i)}$ are constants and $\Delta_{a_i}=|{\vec r}-{\vec a}_i|$ (${\vec r}$ denotes a position vector on ${\mathbb E}^3$ and all ${\vec a}_i$ denote the locations of nuts, which mean zeros of a Killing vector field). Here, let us assume that all point sources ${\vec a}_i$ are on the $z$-axis on ${\mathbb E}^3$, and then $\Delta_{a_i}=\sqrt{r^2+a_i^2+2ra_i\cos\theta}$. The (multi-)black ring solutions on the base space with $\epsilon=0,Q^{(1)}>0,Q^{(i)}=0(i=2,\cdots)$, i.e., a flat space, were constructed by Elvang et.al.~\cite{Elvang} and Gauntlett et.al~\cite{Gauntlett}. The black ring solutions on the self-dual Taub-NUT space with $\epsilon=1,Q^{(1)}>0,Q^{(i)}=0(i=2,\cdots)$ were also constructed by Bena~\cite{Bena}. The base space with $\epsilon=0,\ Q_k^{(1)}=Q_k^{(2)}=a/8,$ $Q_k^{(i)}=0\ (i\ge 3)$ and $\vec{a_1}=-\vec{a_2}=(0,0,a)$ is called the Eguchi-Hanson space. See Appendix \ref{sec:EH} about the transformation of the Eguchi-Hanson space into the Gibbons-Hawking coordinate.
Especially if all point sources of the harmonics $H_k$ are located on the $z$-axis in ${\mathbb E}^3$, i.e.,  $\vec{a_i}=(0,0,-a_i)$, the one-form $\bm\varphi$ takes the form
\begin{eqnarray}
\bm\varphi=\left(\sum_i\frac{Q^{(i)}_k(r\cos\theta+a_i)}{\Delta_{a_i}}\right)d\phi.
\end{eqnarray}
If we assume that all point sources of three harmonic functions $K,L$ and $M$ also exist on $z$-axis, they are given by
\begin{eqnarray}
K&=&k_0+\sum_i\frac{k_i}{\Delta_{R_i}},\\
L&=&l_0+\sum_i\frac{l_i}{\Delta_{R_i}},\\
M&=&m_0+\sum_i\frac{m_i}{\Delta_{R_i}},
\end{eqnarray}
where $\Delta_{R_i}$ are given by
\begin{eqnarray}
& &\Delta_{R_i}:=\sqrt{r^2+R_i^2+2R_ir\cos\theta}.
\end{eqnarray}
Then, the one-forms $\tilde{\bm\omega}=\tilde\omega_\phi d\phi$ and $\tilde{\bm\beta}=\tilde \beta_\phi d\phi$ are computed as

\begin{eqnarray}
\tilde\omega_\phi&=& 
\sum_{i, j}Q^{(i)}m_j\frac{a_i(r\cos\theta+R_j)+r(r+R_j\cos\theta )}{(a_i-R_j)\Delta_{a_i}\Delta_{R_j}}
\nonumber\\
&&+\sum_{i\not=j}\frac{3}{2}k_il_j\frac{R_i(r\cos\theta+R_j)+r(r+R_j\cos\theta )}{(R_i-R_j)\Delta_{R_i}\Delta_{R_j}}
\nonumber\\
&&+
\sum_i\left(\epsilon m_i+\frac{3}{2}k_0l_i-\frac{3}{2}l_0k_i\right)\frac{r\cos\theta+R_i}{\Delta_{R_i}}
\nonumber\\
&&-
\sum_iQ^{(i)}m_0\frac{r\cos\theta+a_i}{\Delta_{a_i}}+C_\omega,
\end{eqnarray}

\begin{eqnarray}
\tilde\beta_\phi=-\sum_ik_i\frac{r\cos\theta+R_i}{\Delta_{R_i}}+C_\beta,
\end{eqnarray}
where we used the result of Appendix \ref{sec:omega}, $C_\omega$ and $C_\beta$ are arbitrary constants. In this article, we study black ring solutions on the Eguchi-Hanson space in the simplest case of $k_i=l_i=m_i=0\ (i\ge 2)$. We put $k_0=0$ so that the metric component does not diverge at the spatial infinity. We can always put $C_\beta=0$ from the freedom of the gauge.

\section{Black ring solutions}\label{sec:solutions}

Here, we determine the parameters in our solutions so that our solutions describe a rotating black ring on the Eguchi-Hanson space.


\subsection{Determination of parameters}

\subsubsection{Absence of Dirac-Misner strings}

To begin with, we consider the condition that Dirac-Misner strings does not exist everywhere outside an event horizon. The existence would yields closed timelike curves if one impose that there is no conical singularity. In our solutions
the conditions $\tilde\omega_\phi(\theta=0)=0$ and $\tilde\omega(\theta=\pi)=0$ assure the absence of Misner strings.  Without loss of generality, we can assume that $R_1\ge 0$. One of the conditions 
$\tilde\omega_\phi(\theta=0)=0$ with $k_0=0$ is computed as follows
\begin{eqnarray}
C_\omega+\frac{m_0a}{8}\left(\frac{a-r}{|a-r|}-1\right)-\frac{3}{2}k_1l_0+\frac{m_1a}{8}\left(\frac{1}{a-R_1}+\frac{a-r}{|a-r|(a+R_1)}\right)=0.\label{eq:omega1}
\end{eqnarray}
The other condition $\tilde\omega(\theta=\pi)=0$ is computed as
\begin{eqnarray}
&&
C_\omega+\frac{m_0a}{8}\left(-\frac{a-r}{|a-r|}+1\right)+\frac{3}{2}k_1l_0\frac{r-R_1}{|r-R_1|}\nonumber\\
 & & -\frac{m_1a}{8}\left(\frac{(a-r)(r-R_1)}{|a-r||r-R_1|(a-R_1)}+\frac{r-R_1}{|r-R_1|(a+R_1)}\right)=0.\label{eq:omega2}
\end{eqnarray}
We see that there are not $C_\omega$, $m_0$ and $m_1$ satisfying Eqs.(\ref{eq:omega1}) and (\ref{eq:omega2}) expect the case of $R_1=0$. Hence, we consider only the case of $R_1=0$, and then the parameters $C_\omega$, $m_0$ and $m_1$  are determined in terms of $k_1$ and $l_0$ as follows
\begin{eqnarray}
C_\omega=0,\quad m_0=-\frac{6k_1l_0}{a},\quad m_1=6k_1l_0.\label{eq:cons}
\end{eqnarray}

\subsubsection{Asymptotic condition}
As mentioned in Introduction, we construct solutions with asymptotically locally Euclidean timeslices. 
%
Therefore, to assure that $g_{tt}\simeq O(1)$ near the spatial infinity $r\to\infty$, we must require $k_0=0$. Under this requirement, 
in the neighborhood of the spatial infinity $r\to\infty$, the metric behaves as follows
\begin{eqnarray}
ds^2&\simeq&\left(-\frac{1}{l_0^2}+\frac{2(4k_1^2+l_1a)}{l_0^3ar}\right)dt^2\nonumber\\
     & &+\left(-\frac{C_\omega}{l_0^2}+\frac{2C_\omega(8k_1^2+2l_1a)-l_0(8k_1^3+3k_1l_1a+m_1a^2/2)\cos\theta}{2l_0^3ar}\right)dtd\phi\nonumber\\
      & &+\biggl(-\frac{6k_1l_0+m_0a}{8l_0^2}+\frac{16k_1^3l_0+3k_1l_0l_1a+4k_1^2m_0a+l_1m_0a^2-l_0m_1a^2/2}{4l_0^3ar}\biggr)dtd\psi\nonumber\\
      & &+\frac{l_0a}{4r}dr^2+\frac{rl_0a}{16}(4d\phi^2+d\psi^2+4\cos\theta d\phi d\psi)+\frac{l_0ar}{4}d\theta^2,
\end{eqnarray}
where we used $R_1=0$. To obtain black ring solutions with the desirable asymptotic structure, we must choose the parameters  $l_0$, $C_\omega$ and $m_0$ as
\begin{eqnarray}
&&l_0=1\\
&&C_\omega=0\\
&&6k_1l_0+m_0a=0,
\end{eqnarray}
where we choose $l_0>0$ so that $g_{\phi\phi}>0$ and $g_{\psi\psi}>0$ for large $r$, which is the requirement for absence of closed timelike curves near the spatial infinity.
It should be noted that this choice of the parameters is consistent with the first and the second equations of Eq. (\ref{eq:cons}) obtained from the requirements for the absence of Dirac Misner stings.



\subsection{Black ring solutions}
Thus, from the requirement of the absence of Dirac-Misner strings and asymptotic condition, we obtain the following metric
\begin{eqnarray}
ds^2&=&-H^{-2}\left[d t+\omega_0\left(\frac{a}{8}d\psi+\varphi_\phi d\phi\right)+\tilde\omega_\phi d\phi\right]^2\nonumber\\
 &&+H\left[H_k(dr^2+r^2d\Omega_{\rm S^2}^2)+H_k^{-1}\left(\frac{a}{8}d\psi+\varphi_\phi d\phi\right)^2\right],
\end{eqnarray}
where $d\Omega_{\rm S^2}^2=d\theta^2+\sin^2\theta d\phi^2$ and  we introduced a new coordinate defined by $\psi:=8\xi/a$. The coordinates $r,\psi,\phi,\theta$ run the ranges
\begin{eqnarray}
r>0,\ 0\le \psi\le 4\pi,\ 0\le \phi\le 2\pi,\ 0\le\theta\le\pi.
\end{eqnarray}
The five functions $H_k,H,\omega_0,\tilde\omega_\phi$ and $\varphi_\phi$ are given by 
\begin{eqnarray}
H_k=\frac{a}{8}\left(\frac{1}{\Delta_a}+\frac{1}{\Delta_{-a}}\right),
\end{eqnarray}
\begin{eqnarray}
H=1+\frac{l_1}{r}+\frac{8k_1^2\Delta_{a}\Delta_{-a}}{ar^2(\Delta_{a}+\Delta_{-a})},
\end{eqnarray}
\begin{eqnarray}
\omega_0=2k_1\left(-\frac{3}{a}+\frac{3}{r}+\frac{6(l_1+r)\Delta_{a}\Delta_{-a}}{ar^2(\Delta_{a}+\Delta_{-a})}+\frac{32k_1^2\Delta_{a}^2\Delta_{-a}^2}{a^2r^3(\Delta_{a}+\Delta_{-a})^2}\right),
\end{eqnarray}
\begin{eqnarray}
\tilde\omega_\phi&=&\frac{3k_1}{4\Delta_{a}\Delta_{-a}}[(r+a)(\Delta_{-a}-\Delta_{a})+((r+a)(\Delta_{-a}+\Delta_{a})-2\Delta_{-a}\Delta_{a})\cos\theta],
\end{eqnarray}
\begin{eqnarray}
\varphi_\phi=\frac{a}{8}\left(\frac{a(\Delta_{-a}-\Delta_{a})+r(\Delta_{a}+\Delta_{-a})\cos\theta}{\Delta_{a}\Delta_{-a}}\right).
\end{eqnarray}
It is noted that our solutions have three independent parameters $l_1,k_1$ and $a$, where $k_1$ and $l_1$ are related to the dipole charge $q$ of the black ring and the total electric charge $Q_e$ by $k_1=-q/2$ and $al_1=4G_5Q_e/(\sqrt{3}\pi)-q^2$, and $a$ is the radius of the $\rm S^2$-bolt on the Eguchi-Hanson space. Furthermore, we impose the following conditions on these parameters
\begin{eqnarray}
\quad k_1<0,\quad l_1>-4k_1.
\end{eqnarray}
As explained later, these are the conditions for the absence of closed timelike curves everywhere outside the event horizon. It should be noted that though we choose the origin of the three-dimensional Euclid space ${\mathbb E}^3$ in the Gibbons-Hawking coordinate in the above metric such that that  $\vec{a_1}=-\vec{a_2}=(0,0,a)$ and $R_1=0$, we do not loose the generality. For example, if we shift the origin so that $\vec{a_1}=0$, we change $\Delta_{-a}$, $\Delta_{a}$ and $r$ into $r$, $\Delta_{2a}$ and $\Delta_{a}$, respectively.


\section{Properties}\label{sec:properties}

\subsection{Asymptotic structure}

Let us introduce a new coordinate defined by $\tilde r^2:=ar$. Then, under the choice of the parameters (\ref{eq:cons}), the asymptotic form of the metric for $r\to\infty$ becomes
\begin{eqnarray}
ds^2&\simeq&\left(-1+\frac{2(4k_1^2+l_1a)}{\tilde r^2}\right)dt^2
\nonumber\\
     & &
-\frac{k_1(3a^2+8k_1^2+3al_1)\cos\theta}{\tilde r^2}dtd\phi
-\frac{k_1(3a^2+8k_1^2+3al_1)}{2\tilde r^2}dtd\psi
\nonumber\\
     & &
+d\tilde r^2+\frac{\tilde r^2}{4}\left[\left(\frac{d\psi}{2}+\cos\theta d\phi\right)^2+d\theta^2+\sin^2\theta d\phi^2\right],\label{eq:asy}
\end{eqnarray}
which means that the spatial infinity is topologically the lens space $L(2;1)=S^3/{\mathbb Z}_2$. The asymptotic form of the metric on $t=$constant surfaces resembles the four-dimensional Euclid space, but they differ from each other in the topology of $r={\rm constant}$ surfaces on timeslices. We can regard $\rm S^3$ and the lens space $L(2;1)=\rm S^3/{\mathbb Z}_2$ as $\rm S^1$ bundle over $\rm S^2$. The difference between these metric appears in the term of $d\psi$ in (\ref{eq:asy}). If $d\psi/2$ in Eq.(\ref{eq:asy}) is replaced by $d\psi$, the topology of $r={\rm constant}$ surfaces is $\rm S^3$, i.e., the timeslices is asymptotically Euclidean. Furthermore, we introduce new angular variables $\tilde\phi=(2\phi+\psi)/4$, $\tilde\psi=(-2\phi+\psi)/4$ and $\Theta=\theta/2$, and then the asymptotic form of the metric can be rewritten as
\begin{eqnarray}
ds^2&\simeq&\left(-1+\frac{2(4k_1^2+l_1a)}{\tilde r^2}\right)dt^2
\nonumber\\
    & &
-\frac{2k_1(a^2+8k_1^2+3al_1)\cos^2\Theta}{\tilde r^2}dtd\tilde\phi
+\frac{2k_1(a^2+8k_1^2+3al_1)\sin^2\Theta}{\tilde r^2}dtd\tilde\psi
\nonumber\\
    & &
+d\tilde r^2+\tilde r^2\left(\cos^2\Theta d\tilde\phi^2+\sin^2\Theta d\tilde\psi^2 +d\Theta^2\right).\label{eq:asymp2}
\end{eqnarray}
{}From its asymptotic form (\ref{eq:asymp2}) of the metric,
the ADM mass and angular momentums of our solutions can be computed as
\begin{eqnarray}
& &{\cal M}_{\rm ADM}=\frac{\sqrt{3}}{2}|Q_e|=\frac{3\pi}{8G_5}(4k_1^2+al_1),\\
& &J_{\tilde\phi}=J_{\tilde\psi}=-\frac{\pi}{16G_5}k_1(a^2+8k_1^2+3al_1).
\end{eqnarray}
{}From the relationship between the mass and the electric charge, we see that the BPS inequality is saturated.
It is worth noting that two angular momentums of our solutions are equal in contrast to the black ring solutions on a flat base space~\cite{Elvang}. The property in that solutions have the same two angular momentum components is similar to that of the BMPV black hole solutions.


\subsection{Near-horizon geometry}
First, let us shift a origin of three-dimensional Euclid space so that $a_1=0$, $a_2=2a$ and $R=a$. Next, we introduce the coordinate $(x,y,\hat\phi,\hat\psi)$ defined by
\begin{eqnarray}
r=-a\frac{x+y}{x-y},\quad \cos\theta=-1+2\frac{1-x^2}{y^2-x^2}=1-2\frac{y^2-1}{y^2-x^2},
\end{eqnarray}
\begin{eqnarray}
\phi=\hat\phi-\hat\psi,\quad \psi=\hat\phi+\hat\psi.
\end{eqnarray}
As seen later, the horizon is located on $y=-\infty$. Furthermore, let us define a new coordinates $(z,\zeta)$ given by
\begin{eqnarray}
z=-\frac{P}{y},\quad x=\cos\zeta,
\end{eqnarray}
where $P$ is a constant with dimension of length. The location of the event horizon corresponds to $z=0$. To see the geometry in the neighborhood of the event horizon, we introduce the following coordinates $(v,\hat\phi',\hat\psi')$
\begin{eqnarray}
&&dt=dv-\left(B_0+\frac{B_1}{z}+\frac{B_2}{z^2}\right)dz,\\
&&d\hat\phi=d\hat\phi'-\left(C_0+\frac{C_1}{z}\right)dz,\\
&&d\hat\psi=d\hat\psi'-\left(C_0+\frac{C_1}{z}\right)dz.
\end{eqnarray}
Here, the five constants $B_0,B_1,B_2,C_0$ and $C_1$
are given by
\begin{eqnarray}
&&B_0=-\frac{-48k_1^4+72al_1k_1^2+3l_1^2k_1^2+12a^2(-8k_1^2+l_1^2)}{8ak_1P\sqrt{3l_1^2-48k_1^2}},\\
&&B_1=\frac{-48k_1^4+12al_1k_1^2-al_1^3}{4ak_1\sqrt{3l_1^2-48k_1^2}},\\
&&B_2=-\frac{k_1\sqrt{3l_1^2-48k_1^2}P}{4a},\\
&&C_0=-\frac{6l_1}{P\sqrt{3l_1^2-48k_1^2}},\\
&&C_1=-\frac{8k_1^2}{aP\sqrt{3l_1^2-48k_1^2}},
\end{eqnarray}
%
%
where the constants $B_2$, $C_1$ and $B_1$ are chosen to cure the divergences $1/z$ in $g_{\hat\psi' z}$, $1/z^2$ and $1/z$ in $g_{zz}$, respectively. $C_0$ and $B_0$ are determined so that $g_{zz}=O(z)$ for $z\to0$.
Then, under the choice of these constants, the metric behaves as
\begin{eqnarray}
ds^2&\simeq&\frac{a^2C_1}{2k_1P}dvdz-\frac{a[(48k_1^4-12ak_1^2l_1+al_1^3)+(-48k_1^4+3k_1^2l_1^2)]\cos\zeta}{16k_1^2P\sqrt{3l_1^2-48k_1^2}}(dzd\hat\phi'+dzd\hat\psi')\nonumber\\
     &  &+ \left(k_1^2\sin^2\zeta+\frac{3(l_1^2-16k_1^2)a^2}{256k_1^2}\right)(d\hat\phi^{\prime 2}+d\hat\psi^{\prime 2})\nonumber\\
     & &+2\left(-k_1^2\sin^2\zeta+\frac{3(l_1^2-16k_1^2)a^2}{256k_1^2}\right)d\hat\phi'd\hat\psi'+k_1^2d\zeta^2,\label{eq:near}
\end{eqnarray}
for $z\to 0$. 
Since each component take the finite value, we see that $(v,\hat\phi',\hat\psi',z,\zeta)$ are good coordinates in the neighborhood of the event horizon. Hence, $z=0$ corresponds to the Killing horizon since the Killing vector $\partial/\partial v$ becomes null there. For $z\to 0$, the induced metric on $v,z=$constant surfaces, i.e., the spatial cross section of the event horizon becomes 
\begin{eqnarray}
ds^2|_{\cal H}\simeq \frac{3(l_1^2-16k_1^2)a^2}{256k_1^2}d\phi_2^2+k_1^2(d\zeta^2+\sin^2\zeta d\phi_1^2),
\end{eqnarray}
where $\phi_1:=\hat\phi'-\hat\psi'=\hat\phi-\hat\psi=\phi$ and $\phi_2:=\hat\phi'+\hat\psi'$. It should be noted that $\partial/\partial\phi_2|_{z=0}=\partial/\partial\psi|_{z=0}$ and $0\le \phi_1\le 2\pi$.
This implies that the spatial topology of the event horizon is $\rm S^1\times \rm S^2$.


\subsection{Absence of closed timelike curves}
In order to cure the existence of CTCs (closed timelike curves) near the horizon $y\to-\infty$, we need assume the parameters $k_1$ and $l_1$ satisfy the inequality
\begin{eqnarray}
l_1>-4k_1.\label{eq:ctc}
\end{eqnarray}
This inequality and $k_1<0\ (l_1>0)$ is also the sufficient condition to avoid any CTCs everywhere outside the event horizon. To see this, it is sufficient to show the $(\phi,\psi)$-part of the metric
\begin{eqnarray}
ds^2|_{(\phi,\psi)} &=& -H^{-2}[\omega_0(d\xi+\varphi_\phi d\phi)+\tilde \omega_\phi d\phi]^2
\nonumber\\
    & &
+H[H_kr^2\sin^2\theta d\phi^2+H_k^{-1}(d\xi+\varphi_\phi d\phi)^2]
\end{eqnarray}
is positive-definite under the condition (\ref{eq:ctc}). This metric is positive-definite if and only if the following two-dimensional matrix is positive-definite
\begin{eqnarray}
M=\left(
\begin{array}{cc}
A& -C\\
-C& B
\end{array}
\right),
\end{eqnarray}
where $A,B$ and $C$ are given by
\begin{eqnarray}
A=H^3H_k-\omega_0^2,\quad B=H^3H_kr^2\sin^2\theta-\tilde\omega_\phi^2,\quad C=\omega_0\tilde \omega_\phi.
\end{eqnarray}
Therefore, noting that $AB-C^2=H^3H_k(AH_k^2r^2\sin^2\theta-\tilde\omega_\phi)$, we obtain the condition
\begin{eqnarray}
M>0 &\Longleftrightarrow& A>0, \quad AB-C^2>0\\
       &\Longleftrightarrow& AH_k^2r^2\sin^2\theta-\tilde\omega_\phi>0.
\end{eqnarray}
As a result, it is enough to prove the $D:=AH_k^2r^2\sin^2\theta-\tilde\omega_\phi >0$. If we fix the parameter $k_1$ and all variables $(r,\theta)$, $D$ becomes the cubic equation with respect to $l_1$. Namely, $D$ takes the form 
\begin{eqnarray}
D(l_1)=D_3l_1^3+D_2l_1^2+D_1l_1^1+D_0,
\end{eqnarray}
where $D_i\ (i=0,1,2,3)$ are functions dependent on $r,\theta$ and $a,k_1$. The explicit expression is given by Appendix \ref{sec:order2}. Since $D_3,D_2$ and $D_1$ are positive for $l_1>0$, $dD/dl_1$ is positive. Hence, $D$ is a monotonically increasing function of $l_1$. Therefore, under the range (\ref{eq:ctc}) of $l_1$,
\begin{eqnarray}
D(l_1)&> &D\left(l_1=-4k_1\right).
\end{eqnarray} 
In turn, if we fix the parameter $l_1$ as $l_1=-4k_1$, $D':=D(-4k_1)$ becomes the quartic equation with respect to $k_1$. 
\begin{eqnarray}
D'=D'_4k_1^4+D'_3k_1^3+D'_2k_2+D'_1k_1+D'_0,
\end{eqnarray}
where $D'_i\ (i=0,\cdots,4)$ are explicitly written in Appendix \ref{sec:ctc}. Since we can show $D'_4\ge0,D'_2\ge 0,D'_0\ge 0$ and $D'_3\le 0,D'_1\le 0$, we see that $D'>0$ for $k_1<0$. Consequently, this fact means $D>0$ under (\ref{eq:ctc}). Therefore, the condition (\ref{eq:ctc}) with $k_1<0$ is the necessary and sufficient for CTCs to vanish outside the event horizon.

\subsection{Regularity}
In this subsection, we show the absence of curvature singularity outside the event horizon. For this end, it is sufficient to investigate the behavior of the metric near the event horizon and the nuts on the $\rm S^2$-bolt, because some components of the metric and the derivatives diverge only at these places and apparently analytic at the other places. First, let us focus on the near-horizon geometry.
In the neighborhood of the horizon $z=0$, the metric behaves as
\begin{eqnarray}
g_{\mu\nu}\simeq g_{\mu\nu}^{(0)}+g_{\mu\nu}^{(1)}z+g_{\mu\nu}^{(2)}z^2+O(z^3),
\end{eqnarray}
where the Greek indices $\mu,\nu$ run $v,\hat\phi',\hat\psi',z,$ and $\zeta$. Each leading term $g_{\mu\nu}^{(0)}$ are given by Eqs.(\ref{eq:near}). Since the explicit expressions of $g_{\mu\nu}^{(1)}$, $g_{\mu\nu}^{(2)}$ are considerably lengthy, we do not written them here. See Appendix \ref{sec:order2} about the concrete expressions.  Next, let us investigate the behavior of the metric near one ${\vec a}_1$ of the nuts. Let us the shift an origin on ${\mathbb E}^3$ so that ${\vec a}_1=0$ and introduce a new coordinate $R:=\sqrt{(a+l_1)r/2}$. Then, each component is analytic at $R=0$. Here, each component of the metric is $C^2$ at least at these places. Therefore, these results assure the absence of a curvature singularity everywhere outside the horizon.

\subsection{Black hole limit}
Finally, we consider the limit of our black ring to the black hole. Setting $l_1=\mu/a$ ($\mu>0:\rm constants$) and taking the limit of $a\to 0$ in our solution yields the following metic
\begin{eqnarray}
ds^2&=&-\left(1+\frac{4k_1^2+\mu}{\tilde r^2}\right)^{-2}\left[dt+\frac{k_1(8k_1^2+3\mu)}{2\tilde r^2}\left(\frac{d\psi}{2}+\cos\theta d\phi\right)\right]^2\nonumber\\
&&+\left(1+\frac{4k_1^2+\mu}{\tilde r^2}\right)\left[d\tilde r^2+\frac{\tilde r^2}{4}d\Omega_{\rm S^2}^2+\frac{\tilde r^2}{4}\left(\frac{d\psi}{2}+\cos\theta d\phi \right)^2\right].\label{eq:bh}
\end{eqnarray}
This is equal to the metric of the BMPV black hole solutions with the mass parameter $m=4k_1^2+\mu$ and the angular momentum parameter $j=k_1(8k_1^2+\mu)$ except that $d\psi$ is replaced with $d\psi/2$. This difference means that they differ in the topology of $\tilde r=\rm constat$, i.e.,  while the BMPV black hole has a squashed $\rm S^3$ horizon, the spatial topology of the black hole horizon in (\ref{eq:bh}) is the squashed lens space $L(2;1)=\rm S^3/{\mathbb Z}_2$. If we do not set $l_1=\mu/a$ and take the limit $a\to 0$, the metric coincides with that of the BMPV black hole with $j=m^{3/2}$, whose horizon topology is the squashed lens space $L(2;1)=\rm S^3/{\mathbb Z}_2$.

\section{Summary and Discussion}\label{sec:summary}

We have constructed a new supersymmetric stationary black ring solution on the Eguchi-Hanson space which has asymptotically locally Euclidean timeslices as solutions of five-dimensional minimal supergravity. We also have investigated the properties of our solutions. Following the classification of the five-dimensional supersymmetric solutions by Gauntlett et.al.~\cite{Gauntlett3}, it is assured that our solutions preserve supersymmetry. We have shown the absence of curvature singularity and CTCs. Our black ring has same two angular momentum components in contrast with the supersymmetric black ring solution on flat base space obtained by Elvang et.al.~\cite{Elvang}, which has two independent angular momentum components. One of the striking features of our black ring is that the $\rm S^1$-direction of the black ring must be along the equator on a $\rm S^2$-bolt to cure Dirac-Misner strings. Furthermore, we have studied the limit of the black ring to a black hole, which coincides with the BMPV black hole with the topology of the squashed lens space $L(2;1)=\rm S^3/{\mathbb Z}_2$.


Finally, we comment on the possibility of the generalization to multi-black ring solutions on the Eguchi-Hanson space. For simplicity, let us concentrate on the case where two point sources ${\vec a_1},{\vec a_2}$ of the harmonic $H_k$ and all $N$ point sources ${\vec R_1},\cdots,{\vec R_N}$ of three harmonics $K,L,M$ (, i.e., $N$ black rings on the $z$-axis) are put on the $z$-axis in the three-dimensional Euclid space in the Gibbons-Hawking coordinate. Whether there exist multi-black ring solutions is essentially determined by the conditions for the absence of Dirac-Misner strings, namely, the existence of the parameters $C,m_0,\cdots, m_N$ satisfying the conditions $\tilde\omega_\phi(\theta=0)=\tilde\omega_\phi(\theta=\pi)=0$. For example, in case of concentric black rings on a flat base space~\cite{Gauntlett}, the number of these independent conditions amounts to $N+1$ since the $z$-axis on the three-dimensional Euclid space in the Gibbons-Hawking coordinate are divided into $N+2$ intervals by a point source ${\vec a_{1}}$ of the harmonics $H_k$ and $N$ point sources ${\vec R_{1}},\cdots,{\vec R_{N}}$. In general, the conditions $ \tilde\omega_\phi(\theta=0)=\tilde\omega_\phi(\theta=\pi)=0$ for each interval give rise to $N+2$ independent equations to $C,m_0,\cdots,m_N$ satisfying all equations. Therefore, since it is assured that there exist these parameters, the configurations of black rings are arbitrary. In contrast, in the case of the Eguchi-Hanson base space, the situation changes, since there are two point sources ${\vec a_1}$ and ${\vec a_2}$ associated with two nut charges $Q^{(1)}_k$ and $Q^{(2)}_k$ respectively, compared with the flat base space, the number of the intervals on the $z-$axis increase by one. Namely, the number of independent equations exceeds that of the parameters. Thus, in general, it is impossible to put black rings at arbitrary positions on the $z$-axis unlike concentric black ring solutions on the flat space. However, it should be noted that if we impose the location of the point sources $R_1,\cdots,R_N$ on the $z$-axis on the reflection symmetry, i.e., if we choose the parameters to be 
\begin{eqnarray}
&&m_1=m_{M+1},\cdots,m_M=m_{2M},\ k_1=k_{M+1},\cdots,k_M=k_{2M},\ l_1=l_{M+1},\cdots,l_M=l_{2M}\nonumber\\
&& R_1=-R_{M+1},\cdots,R_N=-R_{2M}\nonumber
\end{eqnarray}
 if $N$ is even ($N=2M$ for a positive integer $M$),
and 
\begin{eqnarray}
&&m_1=m_{M+1},\cdots,m_M=m_{2M},\ k_1=k_{M+1},\cdots,k_M=k_{2M},\ l_1=l_{M+1},\cdots,l_M=l_{2M},\nonumber\\
&&R_1=-R_{M+1},\cdots,R_N=-R_{2M},\ R_{2M+1}=0\nonumber
\end{eqnarray}
if $N$ is odd ($N=2M+1$), the black rings can be located on arbitrary places on the $z$-axis except ${\vec r}={\vec a_1},{\vec a_2}$ since the number of independent equations coincides with that of the parameters. Especially, in the case of a single black ring, the black ring admits only the configuration such that the $\rm S^1$-direction of the black ring must be along the equator on a $\rm S^2$-bolt. We leave the detail analysis on the multi-black ring solution on the Eguchi-Hanson space for the future.

\section*{Acknowledgements}
This work is supported by the Grant-in-Aid
for Scientific Research No.14540275 and No.13135208.

\appendix
\section{Eguchi-Hanson space}\label{sec:EH}

The metric of the Eguchi-Hanson space is given by
\begin{eqnarray}
&& ds^2_{\rm EH} = \left( 1- \frac{a^4}{\bar r^4}\right) ^{-1} d\bar r^2 
        + \frac{\bar r^2}{4} \left[ \left( 1-\frac{a^4}{\bar r^4} \right) 
          \left( d\bar\psi+\cos\bar\theta d\bar\phi \right)^2
        + d\bar\theta^2+\sin^2\bar\theta d\bar\phi^2 \right],
\label{metric_EH}
\end{eqnarray}
where $a$ are a constant, $0\leq \bar\theta\leq \pi,~ 
0\leq \bar\phi\leq 2\pi/$ 
and $0\leq \bar\psi\leq 2\pi$. 
The Eguchi-Hanson space has an S$^2$-bolt at $r=a$, 
where the Killing vector field $\partial/\partial \bar \psi$ vanishes. 

In order to clarify the relationship between the Gibbons-Hawking coordinate and the metric (\ref{metric_EH}), we introduce the coordinates as follows
\cite{Prasad},
\begin{eqnarray}
& & r=a\sqrt{\frac{\bar r^4}{a^4}-\sin^2\bar\theta},\quad 
\tan\theta=\sqrt{1-\frac{a^4}{\bar r^4}}\tan\bar\theta,\quad 
\phi=\bar\psi,\quad 
\psi=2\bar\phi. \notag \\
& &
( 0\leq \theta\leq \pi,~~ 
  0\leq \phi\leq 2\pi,~~
  0\leq \psi\leq 4\pi )
\end{eqnarray}
Then, the metric takes the form of 
\begin{eqnarray}
& & ds^2_{\rm EH} = V ^{-1} (r,\theta) 
               \left[ dr^2 + r^2 \left( d\theta^2 + \sin^2\theta d\phi^2 \right) \right]
               + V(r,\theta) \left( \frac{a}{8} d\psi+\varphi_\phi d\phi \right)^2,
\label{metric_GH}
\\
& & V^{-1} (r,\theta) = \frac{a/8}{|{\bm r}-{\bm r}_1|}
            +\frac{a/8}{|{\bm r}-{\bm r}_2|},\\ 
& &\varphi_{\phi}(r, \theta)
= \frac{a}{8} 
        \left(
  \frac{r \cos\theta - a}{\sqrt{r^2 + a^2 - 2 a r \cos\theta}}
+ \frac{r \cos\theta + a}{\sqrt{r^2 + a^2 + 2 a r \cos\theta}}
\right),
\end{eqnarray}
where ${\bm r}=(x,y,z)$ is the position vector on 
the three-dimensional Euclid space and ${\bm r}_1=(0,0,a)$, 
${\bm r}_2=(0,0,-a)$. 
The metric (\ref{metric_GH}) is the  
Gibbons-Hawking two-center form of the Eguchi-Hanson 
space\cite{Prasad,Eguchi,GH}. 
It is manifest in the coordinate that the space has 
two nut singularities at $\bm r= \bm r_j$ 
where the Killing vector field $\partial /\partial \psi$ vanishes.

\section{Solving $\omega$}\label{sec:omega}
The solutions of the following equations
\begin{eqnarray}
*d{\bm\omega}=\frac{1}{\Delta_{R_i}}d\left(\frac{1}{\Delta_{R_i}}\right)-\frac{1}{\Delta_{R_j}}d\left(\frac{1}{\Delta_{R_i}}\right)
\end{eqnarray}
and
\begin{eqnarray}
*d{\bm\omega}=d\left(\frac{1}{\Delta_{R_i}}\right)
\end{eqnarray}
are given by
\begin{eqnarray}
\bm\omega=\frac{R_i(r\cos\theta+R_j)+r(r+R_j\cos\theta)}{(R_i-R_j)\Delta_{R_i}\Delta_{R_j}}d\phi,
\end{eqnarray}
and
\begin{eqnarray}
\bm\omega=\frac{r\cos\theta+R_i}{\Delta_{R_i}}d\phi,
\end{eqnarray}
respectively.

\section{CTCs}\label{sec:ctc}
If we fix the parameter $k_1$ and all variables, $D$ becomes the cubic equation with respect to $l_1$, whose form is given by
\begin{eqnarray}
D(l_1)=\sum_{n=0}^3D_nl_1^n,
\end{eqnarray}
where $D_n\ (n=0,\cdots,3)$ are given by 
\begin{eqnarray}
D_3&=&\frac{a(\Delta_{a_1}+\Delta_{a_2})\sin^2\theta}{8r^3\Delta_{a_1}\Delta_{a_2}},\\
D_2&=&\frac{3r^2(2k_1^2\Delta_{a_1}^2\Delta_{a_2}^2+a\Delta_{a_1}\Delta_{a_2}(\Delta_{a_1}+\Delta_{a_2})r^2)\sin^2\theta}{8r^4\Delta_{a_1}^2\Delta_{a_2}^2},\\
D_1&=&\frac{3r\sin^2\theta}{2\Delta_{a_1}\Delta_{a_2}}
\bigg[
k_1^2\frac{\Delta_{a_1}\Delta_{a_2}}{r^2}+\frac{a(\Delta_{a_1}+\Delta_{a_2})}{4}
\\ &&
+\frac{(ar^2-24k_1^2(a-r))(\Delta_{a_1}+\Delta_{a_2})}{r^2}
\bigg],\\
D_0&=&-\frac{9k_1^2}{16\Delta_{a_1}^2\Delta_{a_2}^2}\left((a+r)(\Delta_{a_2}-\Delta_{a_1})+((a+r)\Delta_{a_1}+(a+r-2\Delta_{a_1})\Delta_{a_2})\cos\theta\right)^2\nonumber\\
                                 & &+\frac{3k_1^2}{4}\sin^2\theta\left(1-\frac{16k_1^2}{r^2}+\frac{16k_1^2}{ar}\right)+\frac{(\Delta_{a_1}+\Delta_{a_2})\sin^2\theta}{4\Delta_{a_1}\Delta_{a_2}}\left(9k_1^2(r+a)-\frac{ar^2}{2}\right)\nonumber\\
                                 & &+\frac{9k_1^2(\Delta_{a_1}+\Delta_{a_2})^2}{16\Delta_{a_1}^2\Delta_{a_2}^2}(r+a)^2.                                 
\end{eqnarray}
It should be noted that $D_3,D_2$ and $D_1$ are positive. Therefore, $D(l_1)$ is monotonically increasing function in the case of $l_1>0$. As a result, $D>D(l_1=-4k_1)$. Next, let us write $D'$ as a function of $k_1$ as follows
\begin{eqnarray}
D':=D(l_1=-4k_1)=\sum_{n=0}^4D'_nk_1^n,
\end{eqnarray}
where $D'_n\ (n=0,\cdots,4)$ are given by
\begin{eqnarray}
D'_4&=&\frac{192\Delta_{a_1}^2\Delta_{a_2}^2\sin^2\theta}{ra},\\
D'_3&=&-72(\Delta_{a_1}+\Delta_{a_2})\Delta_{a_1}\Delta_{a_2}\left[1+\cos2 \theta+\frac{a}{r}(1+\cos2\theta) \right],\\
D'_2&=&\frac{3}{2}\Biggl[-2\Delta_{a_1}^2\Delta_{a_2}^2+6r\Delta_{a_1}\Delta_{a_2}(\Delta_{a_1}+\Delta_{a_2})-3(r^2+a^2)(\Delta_{a_1}^2+\Delta_{a_2}^2)\nonumber\\
& &-3ar(\Delta_{a_1}-\Delta_{a_1})^2+8a\Delta_{a_1}\Delta_{a_2}(\Delta_{a_1}+\Delta_{a_2})\nonumber\\
& &+3(r+a)(\Delta_{a_1}-\Delta_{a_2})((a+r)\Delta_{a_1}+(a+r-2\Delta_{a_1})\Delta_{a_2})\cos\theta\nonumber\\
& &-(4\Delta_{a_1}^2\Delta_{a_2}^2+a(\Delta_{a_1}+\Delta_{a_2})(2\Delta_{a_1}\Delta_{a_2}+3r(\Delta_{a_1}+\Delta_{a_2}))\cos2\theta\Biggr],\\
D'_1&=&-6ar\Delta_{a_1}\Delta_{a_2}(\Delta_{a_1}+\Delta_{a_2})\sin^2\theta,\\
D'_0&=&\frac{ar^2}{2}\Delta_{a_1}\Delta_{a_2}(\Delta_{a_1}+\Delta_{a_2})\sin^2\theta.
\end{eqnarray}
{}From this, we find $D'_4> 0$, $D'_0>0$, $D'_0>0$, $D'_3<0$ and $D'_1<0$. Therefore, for $k_1<0$, $D'$ is positive.

\section{Near-Horizon geometry}\label{sec:order2}
Near the event horizon, in the coordinate $(v,\hat\phi',\hat\psi',z,\zeta)$, the components of the metric behaves as
\begin{eqnarray}
g_{\mu\nu}\simeq g_{\mu\nu}^{(0)}+g_{\mu\nu}^{(1)}z+g_{\mu\nu}^{(2)}z^2+O(z^3),
\end{eqnarray}
where $g_{\mu\nu}^{(0)}$ is given by (\ref{eq:near}). The non-zero components of $g_{\mu\nu}^{(1)}$ and $g_{\mu\nu}^{(2)}$ are computed as follows
\begin{eqnarray}
&&g_{vv}^{(1)}=0,\\
&&g_{vv}^{(2)}=0,
\end{eqnarray}
\begin{eqnarray}
&&g_{vz}^{(1)}=-\frac{2a(al_1-2k_1^2\cos\zeta)}{k_1LP^2},\\
&&g_{vz}^{(2)}=\frac{a(8a^2+al_1\cos\zeta-4k_1^2\cos^2\zeta)}{k_1LP^3},
\end{eqnarray}
\begin{eqnarray}
&&g_{v\phi}^{(1)}=-\frac{a^2}{4k_1P},\\
&&g_{v\phi}^{(2)}=\frac{a^2(al_1+4k_1^2\cos\zeta)}{16k_1^3P^2},
\end{eqnarray}
\begin{eqnarray}
g_{z\phi}^{(1)}&=&-\frac{a}{64Lk_1^4P^2}\biggl[-192k_1^6-(384a^2+96al_1-12l_1^2)k_1^4+a^2l_1^4\nonumber\\
& &+2k_1^2(96k_1^4+24al_1k_1^2-al_1^3)\cos\zeta
               +(-96k_1^6+6l_1^2k_1^4)\cos2\zeta\biggr]\nonumber\\
g_{z\phi}^{(2)}&=&-\frac{a}{128Lk_1^4P^3}\biggl[2(-96k_1^6+(384a^2+24al_1)k_1^4+(192a^3l_1-al_1^3)k_1^2-10a^3l_1^3))\nonumber\\
               & &+2(-336k_1^6+(768a^2+48al_1-15l_1^2)k_1^4-60a^2l_1^2k_1^2+a^2l_1^4)\cos\zeta\nonumber\\
               & &+(-192k_1^6+5al_1^3k_1^2)\cos2\zeta+2(624k_1^6-3l_1^2k_1^4)\cos3\zeta  \biggr],             
\end{eqnarray}

\begin{eqnarray}
g_{z\psi}^{(1)}&=&-\frac{a}{64Lk_1^4P^2}\biggl[-192k_1^6-(384a^2+96al_1-12l_1^2)k_1^4+a^2l_1^4\nonumber\\
& &+2k_1^2(96k_1^4+24al_1k_1^2-al_1^3)\cos\zeta
               +(-96k_1^6+6l_1^2k_1^4)\cos2\zeta\biggr]\nonumber\\
g_{z\psi}^{(2)}&=&-\frac{a}{128Lk_1^4P^3}\biggl[2(-966k_1^6+(384a^2+24al_1)k_1^4+(192a^3l_1-al_1^3)k_1^2-10a^3l_1^3))\nonumber\\
               & &+2(-336k_1^6+(768a^2+48al_1-15l_1^2)k_1^4-60a^2l_1^2k_1^2+a^2l_1^4)\cos\zeta\nonumber\\
               & &+(-192k_1^6+5al_1^3k_1^2)\cos2\zeta-6(176k_1^6+l_1^2k_1^4)\cos3\zeta  \biggr],             
\end{eqnarray}

\begin{eqnarray}
g_{zz}^{(1)}&=&\frac{1}{4L^2k_1^2P^3}\biggl[al_1(288k_1^4-(17l_1^2+284a^2)k_1^2+24a^2l_1^2+(576k_1^4-33l_1^2k_1^2)\cos2\zeta)\nonumber\\
           &&+(-1536a^2k_1^4-192k_1^6+192a^2l_1^2k_1^2+12l_1^2k_1^4-6a^2l_1^4)\cos\zeta\biggr],\\
g_{zz}^{(2)}&=&-\frac{1}{16k_1^6L^2P^4}\biggl[1152k_1^{10}+(14208a^2-72l_1^2)k_1^8+(2048a^4-1320a^2l_1^2)k_1^6\nonumber\\
           & &+(384a^4l_1^2+32a^2l_1^4)k_1^4-48a^4l_1^4k_1^2+a^4l_1^6 \nonumber\\
           & &+12a l_1k_1^2(-680k_1^6+(41l_1^2+320a^2)k_1^4-36a^2l_1^2k_1^2+a^2l_1^4)\cos\zeta\nonumber\\
           & &+24k_1^4(16k_1^6+(688a^2-l_1^2)k_1^4-65a^2l_1^2k_1^2+2a^2l_1^4)\cos2\zeta\nonumber\\
           & &+al_1k_1^6(-6624k_1^2+384l_1^2)\cos3\zeta \biggr],
\end{eqnarray}

\begin{eqnarray}
g_{\zeta\zeta}^{(1)}&=&\frac{al_1-4k_1^4\cos\zeta}{2P},\\
g_{\zeta\zeta}^{(2)}&=&\frac{2a^2+2k_1^2-3al_1\cos\zeta+2k_1^2\cos\zeta}{2P^2},
\end{eqnarray}

\begin{eqnarray}
g_{\psi\psi}^{(1)}&=&\frac{1}{256k_1^4P}\biggl[(96a^2+64al_1)k_1^4+24a^3l_1k_1^2-a^3l_1^3-128k_1^6\cos\zeta\nonumber\\
                  & &-64al_1k_1^4\cos2\zeta+128k_1^6\cos3\zeta\biggr],\\
g_{\psi\psi}^{(2)}&=&\frac{1}{1024k_1^6P^2}\biggl[-128k_1^8+512a^2k_1^6-96a^3l_1k_1^4-24a^4l_1^2k_1^2+a^4l_1^4\nonumber\\
                  & &-4ak_1^2(96(a+l_1)k_1^4+24l_1a^2k_1^2-a^2l_1^3)\cos\zeta+512k_1^6(k_1^2-a^2)\cos2\zeta\nonumber\\
                  & &+384al_1 k_1^6\cos3\zeta-384k_1^8\cos4\zeta\biggr],
\end{eqnarray}

\begin{eqnarray}
g_{\phi\phi}^{(1)}&=&\frac{1}{256k_1^4P}\biggl[(96a^2+64al_1)k_1^4+24a^3l_1k_1^2-a^3l_1^3-128k_1^6\cos\zeta\nonumber\\
                  & &-64al_1k_1^4\cos2\zeta+128k_1^6\cos3\zeta\biggr],\\
g_{\phi\phi}^{(2)}&=&\frac{1}{1024k_1^6P^2}\biggl[-128k_1^8+512a^2k_1^6-96a^3l_1k_1^4-24a^4l_1^2k_1^2+a^4l_1^4\nonumber\\
                  & &-4ak_1^2(96(a+l_1)k_1^4+24l_1a^2k_1^2-a^2l_1^3)\cos\zeta+512k_1^6(k_1^2-a^2)\cos2\zeta\nonumber\\
                  & &+384al_1 k_1^6\cos3\zeta-384k_1^8\cos4\zeta\biggr],
\end{eqnarray}

\begin{eqnarray}
g_{\phi\psi}^{(1)}&=&\frac{1}{256k_1^4P}\biggl[(96a^2-64al_1)k_1^4+24a^3l_1k_1^2-a^3l_1^3+128k_1^6\cos\zeta\nonumber\\
                  & &+64al_1k_1^4\cos2\zeta-128k_1^6\cos3\zeta\biggr],\\
g_{\phi\phi}^{(2)}&=&\frac{1}{1024k_1^6P^2}\biggl[128k_1^8-512a^2k_1^6-96a^3l_1k_1^4-24a^4l_1^2k_1^2+a^4l_1^4\nonumber\\
                  & &-4ak_1^2(96(a-l_1)k_1^4+24l_1a^2k_1^2-a^2l_1^3)\cos\zeta-512k_1^6(k_1^2-a^2)\cos2\zeta\nonumber\\
                  & &-384al_1 k_1^6\cos3\zeta+384k_1^8\cos4\zeta\biggr],
\end{eqnarray}
where $L:=\sqrt{3l_1^2-48k_1^2}$.

\end{document}